\documentclass[a4paper,11pt]{article}
\pdfoutput=1 

\usepackage{jcappub} 
\usepackage{bm}
\usepackage[table,xcdraw]{xcolor}
\usepackage{hyperref}
\usepackage{graphicx}
\usepackage{cleveref}
\usepackage[T1]{fontenc} 
\usepackage{longtable}
\usepackage[tableposition=t]{caption}
\usepackage{booktabs}
\usepackage{rotating}
\usepackage{enumitem}
\usepackage{float}

\newcommand{\hMpc}{\ h^{-1}\text{Mpc}}

\newcommand{\rpar}{r_\parallel}

\newcommand{\lya}{Ly$\alpha$}

\newcommand{\lrf}{\lambda_{{\rm rest}}}
\newcommand{\apar}{\alpha_\parallel}
\newcommand{\aperp}{\alpha_\perp}

\newcommand{\blya}{b_{\rm Ly\alpha}}
\newcommand{\betalya}{\beta_{\rm Ly\alpha}}
\newcommand{\Hone}{{\rm HI}}
\newcommand{\zl}{z_\lambda}
\newcommand{\Fzl}{\overline{F}(z_\lambda)}
\newcommand{\rperp}{r_\perp}

\newcommand{\Lql}{\Lambda_{q\lambda}}
\newcommand{\Lqlp}{\Lambda_{q\lambda^\prime}}

\newcommand{\nql}{n_{q\lambda}}
\newcommand{\nqlp}{n_{q\lambda^\prime}}

\newcommand{\dtql}{\delta^t_{q\lambda}}
\newcommand{\dmql}{\delta^m_{q\lambda}}
\newcommand{\dmqlp}{\delta^m_{q\lambda^\prime}}
\newcommand{\dtpql}{\delta^{tp}_{q\lambda}}
\newcommand{\dtqlp}{\delta^{t}_{q\lambda^\prime}}
\newcommand{\dmpql}{\delta^{mp}_{q\lambda}}
\newcommand{\dmpqplp}{\delta^{mp}_{q^\prime\lambda^\prime}}
\newcommand{\dmqplp}{\delta^{m}_{q^\prime\lambda^\prime}}

\newcommand{\eql}{\epsilon_{q\lambda}}

\newcommand{\Wq}{W_{q}}
\newcommand{\wql}{w_{q\lambda}}
\newcommand{\wqplp}{w_{q^\prime\lambda^\prime}}

\newcommand{\wqlp}{w_{q\lambda^\prime}}

\newcommand{\fql}{f_{q\lambda}}

\newcommand{\Cql}{C_{q\lambda}}

\newcommand{\Tql}{T_{\lambda z_q}}

\newcommand{\dtqplp}{\delta^t_{q^\prime\lambda^\prime}}

\newcommand{\lp}{\lambda^\prime}
\newcommand{\lpp}{\lambda^{\prime\prime}}
\newcommand{\lppp}{\lambda^{\prime\prime\prime}}

\usepackage{amsmath}
\let\oldAA\AA
\renewcommand{\AA}{\text{\normalfont\oldAA}}

\title{The effects of continuum fitting on Lyman-$\alpha$ forest correlations}

\author[1]{Nicolas Busca,}
\author[2,*]{James Rich,\note{Corresponding author.}}
\author[3]{Julian Bautista,}
\author[4]{Andrei~Cuceu,}
\author[5]{Andreu Font-Ribera,}
\author[4]{Julien Guy,}
\author[2,6]{Hiram K. Herrera-Alcantar,}
\author[1]{Julianna Stermer,}
\author[1]{Christophe Balland,}
\author[4]{J.~Aguilar,}
\author[7]{S.~Ahlen,}
\author[8,9]{D.~Bianchi,}
\author[10]{D.~Brooks,}
\author[4]{T.~Claybaugh,}
\author[11]{A.~de la Macorra,}
\author[10]{P.~Doel,}
\author[4,12]{S.~Ferraro,}
\author[13,14]{J.~E.~Forero-Romero,}
\author[15,16,17]{E.~Gaztañaga,}
\author[5]{C.~Gordon,}
\author[18]{G.~Gutierrez,}
\author[19]{M.~Ishak,}
\author[20]{R.~Kehoe,}
\author[21]{D.~Kirkby,}
\author[4]{A.~Kremin,}
\author[4]{M.~Landriau,}
\author[1]{L.~Le~Guillou,}
\author[2]{C.~Magneville,}
\author[22,23]{P.~Martini,}
\author[5,24]{R.~Miquel,}
\author[16]{S.~Nadathur,}
\author[2,4]{N.~Palanque-Delabrouille,}
\author[25]{F.~Prada,}
\author[26]{I.~P\'erez-R\`afols,}
\author[27]{C.~Ravoux,}
\author[28]{G.~Rossi,}
\author[29]{E.~Sanchez,}
\author[4]{D.~Schlegel,}
\author[30]{H.~Seo,}
\author[4]{J.~Silber,}
\author[31]{D.~Sprayberry,}
\author[32]{G.~Tarl\'{e},}
\author[31]{B.~A.~Weaver,}
\author[4]{R.~Zhou,}
\author[33]{H.~Zou}

\affiliation[1]{Sorbonne Universit\'e, CNRS/IN2P3, Laboratoire de Physique Nucl\'eaire et de Hautes Energies, LPNHE, 4 Place Jussieu, F-75252 Paris, France}
\emailAdd{james.rich@cea.fr}
\affiliation[2]{CEA, IRFU, Universit\'e Paris Saclay, F-91191 Gif-sur-Yvette, France}
\affiliation[3]{Aix Marseille Universit\'e, CNRS/IN2P3, CPPM, Marseille, France}
\affiliation[4]{Lawrence Berkeley National Laboratory, 1 Cyclotron Road, Berkeley, CA 94720, USA}
\affiliation[5]{Institut de F\'{i}sica d’Altes Energies (IFAE), The Barcelona Institute of Science and Technology, Campus UAB, 08193 Bellaterra Barcelona, Spain}
\affiliation[6]{Institut d'Astrophysique de Paris. 98 bis boulevard Arago. 75014 Paris, France}
\affiliation[7]{Department of Physics, Boston University, 590 Commonwealth Avenue, Boston, MA 02215 USA}
\affiliation[8]{Dipartimento di Fisica ``Aldo Pontremoli'', Universit\`a degli Studi di Milano, Via Celoria 16, I-20133 Milano, Italy}
\affiliation[9]{INAF-Osservatorio Astronomico di Brera, Via Brera 28, 20122 Milano, Italy}
\affiliation[10]{Department of Physics \& Astronomy, University College London, Gower Street, London, WC1E 6BT, UK}
\affiliation[11]{Instituto de F\'{\i}sica, Universidad Nacional Aut\'{o}noma de M\'{e}xico,  Circuito de la Investigaci\'{o}n Cient\'{\i}fica, Ciudad Universitaria, Cd. de M\'{e}xico  C.~P.~04510,  M\'{e}xico}
\affiliation[12]{University of California, Berkeley, 110 Sproul Hall \#5800 Berkeley, CA 94720, USA}
\affiliation[13]{Departamento de F\'isica, Universidad de los Andes, Cra. 1 No. 18A-10, Edificio Ip, CP 111711, Bogot\'a, Colombia}
\affiliation[14]{Observatorio Astron\'omico, Universidad de los Andes, Cra. 1 No. 18A-10, Edificio H, CP 111711 Bogot\'a, Colombia}
\affiliation[15]{Institut d'Estudis Espacials de Catalunya (IEEC), c/ Esteve Terradas 1, Edifici RDIT, Campus PMT-UPC, 08860 Castelldefels, Spain}
\affiliation[16]{Institute of Cosmology and Gravitation, University of Portsmouth, Dennis Sciama Building, Portsmouth, PO1 3FX, UK}
\affiliation[17]{Institute of Space Sciences, ICE-CSIC, Campus UAB, Carrer de Can Magrans s/n, 08913 Bellaterra, Barcelona, Spain}
\affiliation[18]{Fermi National Accelerator Laboratory, PO Box 500, Batavia, IL 60510, USA}
\affiliation[19]{Department of Physics, The University of Texas at Dallas, 800 W. Campbell Rd., Richardson, TX 75080, USA}
\affiliation[20]{Department of Physics, Southern Methodist University, 3215 Daniel Avenue, Dallas, TX 75275, USA}
\affiliation[21]{Department of Physics and Astronomy, University of California, Irvine, 92697, USA}
\affiliation[22]{Center for Cosmology and AstroParticle Physics, The Ohio State University, 191 West Woodruff Avenue, Columbus, OH 43210, USA}
\affiliation[23]{Department of Astronomy, The Ohio State University, 4055 McPherson Laboratory, 140 W 18th Avenue, Columbus, OH 43210, USA}
\affiliation[24]{Instituci\'{o} Catalana de Recerca i Estudis Avan\c{c}ats, Passeig de Llu\'{\i}s Companys, 23, 08010 Barcelona, Spain}
\affiliation[25]{Instituto de Astrof\'{i}sica de Andaluc\'{i}a (CSIC), Glorieta de la Astronom\'{i}a, s/n, E-18008 Granada, Spain}
\affiliation[26]{Departament de F\'isica, EEBE, Universitat Polit\`ecnica de Catalunya, c/Eduard Maristany 10, 08930 Barcelona, Spain}
\affiliation[27]{Universit\'{e} Clermont-Auvergne, CNRS, LPCA, 63000 Clermont-Ferrand, France}
\affiliation[28]{Department of Physics and Astronomy, Sejong University, 209 Neungdong-ro, Gwangjin-gu, Seoul 05006, Republic of Korea}
\affiliation[29]{CIEMAT, Avenida Complutense 40, E-28040 Madrid, Spain}
\affiliation[30]{Department of Physics \& Astronomy, Ohio University, 139 University Terrace, Athens, OH 45701, USA}
\affiliation[31]{NSF NOIRLab, 950 N. Cherry Ave., Tucson, AZ 85719, USA}
\affiliation[32]{University of Michigan, 500 S. State Street, Ann Arbor, MI 48109, USA}
\affiliation[33]{National Astronomical Observatories, Chinese Academy of Sciences, A20 Datun Road, Chaoyang District, Beijing, 100101, P.~R.~China}

\abstract{
Correlations of fluctuations of the flux in  Lyman-$\alpha$ forests of high-redshift quasars 
have been observed by the Baryonic Acoustic Oscillation Spectroscopy Survey (BOSS)  and the Dark Energy Spectroscopy Instrument (DESI) survey 
where they have revealed the effects of baryon acoustic oscillations (BAO).
In order to fit the correlation functions to a physical model and thereby constrain cosmological parameters, it is necessary to take into account the effects of fitting the observed spectra to a template about which the fluctuations are measured.
In this paper we use mock spectra to test the distortion matrix technique that has been used 
since the final BOSS data release
to appropriately distort the models.
We show that while percent-level effects on the  derived forest bias parameters may be present, the technique works sufficiently well that the
determination of the BAO peak position is not affected at the percent level.
We introduce modifications in the technique used by DESI that were not in the original
applications 
and suggest further possibilities for improvements.
}

\begin{document}
\maketitle
\flushbottom

\section{Introduction}

The \lya~forest is a series of absorption features in the
spectra of high-redshift quasars 
due to the presence of neutral hydrogen (\Hone)
in the intergalactic medium (IGM).
Correlations in the wavelength-dependent absorption in
forests are related to the correlations in the
underlying matter density.
The Baryon Oscillation Spectroscopy Survey (BOSS)
\cite{2013AJ....145...10D}, 
the extended Baryon Oscillation Spectroscopy Survey (eBOSS) \citep{2016AJ....151...44D},
and the ongoing Dark Energy Spectroscopic Instrument (DESI) project \citep{Snowmass2013.Levi,2016arXiv161100036D,DESI2016b.Instr}\footnote{
described in detail in 
\citep{DESI:2022xcl,
FocalPlane.Silber.2023,
FiberSystem.Poppett.2024,
Corrector.Miller.2023, 
Spectro.Pipeline.Guy.2023,
SurveyOps.Schlafly.2023,
DESI2023b.KP1.EDR,DESI2023a.KP1.SV}.
}
have obtained a sufficient number of spectra to use these correlations to study cosmological
large-scale structure (LSS).
The auto-correlations 
between neighboring forests were first observed by \citep{2011SlosarBOSS}
and 
 the cross-correlations
between forests and quasars were first seen by \citep{Font+13}.
The imprint of baryon acoustic oscillations (BAO)
were then observed 
\citep{2013BuscaBOSS,2013SlosarDR9,Font+14,2015DelubacDR11,2017BautistaDR12,dmdB+17,2019deSainteAgatheDR14,2019BlomqvistDR14,dMdB+20,DESIyr1lyabao,DESI-yr3lyabao}.
The cosmological constraints thus derived from BOSS-eBOSS were presented in \citep{eBOSS21}
and those from the DESI Data Release 1 (DR1) \citep{DesiDR1_2025arXiv250314745D} were presented in \citep{DESIyr1cosmo,Desi_fullshape_2024arXiv241112022D} and those from Data Release 2 (DR2) in \citep{DESI-yr3lyabao}.

\begin{figure}
\centering
\includegraphics[width=\textwidth]{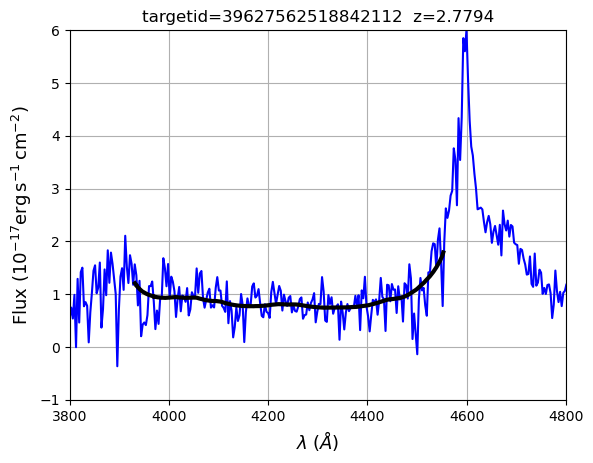}
\caption{
    A high signal-to-noise  DESI quasar spectrum
    as a function of observed wavelength.
    The black curve is
    the estimated continuum, $\Fzl C_{q}(\lrf)$,
    over the restframe wavelength range $1040<\lrf<1200\AA$ used in this study.
      }
\label{fig:oneforest}
\end{figure}

The flux spectrum of a DESI quasar is shown in Fig. \ref{fig:oneforest}.
The flux-transmission field is found
by comparing the observed flux at each wavelength with the expected flux assuming
no fluctuations, the solid black line in the figure.
Unfortunately, the expected flux is not known  apriori but must
be calculated by fitting the form of the spectrum in the forest
region, the so-called continuum fitting.
This mixes the
absorption of different wavelengths, causing a significant
distortion of the correlations between different forests,
complicating the determination of cosmological parameters.\footnote{Attempts have also been made to derive the forest continuum by fitting only the spectrum on the red side of \lya~emission, but these fit have not yet been sufficiently accurate to avoid fitting in the forest region\citep{Lee2013,Turner_2024ApJ...976..143T}.}

Early attempts to account for this effect were presented in \citep{2011SlosarBOSS}, \citep{Font+12dla}, \citep{2013SlosarDR9}, and \citep{Blomqvist2015}.
The authors of \citep{2011SlosarBOSS} used a simplified model for the effect of continuum fitting on the flux-transmission field and derived modifications to the correlation function. 
In \citep{Font+12dla},  a similar approach was used to model the impact in the cross-correlation of galaxies and the Lyman alpha forest. An alternative approach was
later developed in \citep{2013SlosarDR9} by moving the effect of continuum fitting from the model to the covariance matrix, such that linear combinations of data points  that are most impacted by continuum fitting can be marginalized out when fitting cosmological models. In parallel, the authors of \citep{Blomqvist2015} also attempt a principled approach in power-spectrum space and encapsulate the effect of continuum fitting in a multiplicative function with free parameters that can be fit to the data together with the cosmological model. 

Starting with the  BOSS analysis of SDSS DR12 \citep{2017BautistaDR12}, all analyses have used the "distortion matrix"  formalism  described in this paper.
Inspired by \citep{2011SlosarBOSS} and \citep{2013SlosarDR9},
this approach aims to distort the model correlation function used to fit the data in the same way as the continuum fitting distorts the measured correlation function.
In addition to the \lya~forest auto-correlation and the quasar-forest cross-correlation, the technique has also been used in the study of correlations between forests and Damped Ly$\alpha$ systems \citep{dlaforestcor_2018MNRAS.473.3019P}, between forests and the circumgalactic medium \citep{galaxyforestcorr_2023MNRAS.524.1464P}, and correlations involving metalic absorbers \citep{Mg_2019ApJ...878...47D}.

 The purpose of this paper is to describe the distortion of the correlation functions due
to continuum fitting and how physical models of the undistorted correlations can be given the
same distortion using the distortion matrix.
We then test the distortion matrix using mock data sets where both the measured and
true fields are known.
The test consists of verifying that the correlation functions of the true fields transformed by the distortion matrix reproduces the measured correlation functions.

The distortion matrix has been already tested indirectly in \lya~studies of BAO and 
of the Alcock-Paczynski (AP) effect  \citep{Cuceu_2023PhRvL.130s1003C} where
it was shown that the complete analysis chain (including the distortion matrix) of the mock data sets returned the BAO and AP parameters that
were used to create the mocks.
Here we present direct tests that isolate the effects of the distortion matrix on the 
full shape of the correlation functions.  We can therefore study effects on the measurement of forest bias parameters.

For the tests described in this paper we use the "LyaCoLoRe" mocks \citep{Farr+19,colore_2022JCAP...05..002R,Herrera_2024} developed for  BOSS and DESI.
These mocks are generated using Gaussian-random fields to which are applied the Fluctuating Gunn-Peterson Approximation \citep{FGPA_weinberg1998cosmologylymanalphaforest}.
Absorption due to metals and damped Ly$\alpha$ systems is not included since these complications are not expected to have an impact on the continuum-fitting distortions.
The  mock sets reproduce the DESI geometry and spectral resolution.
The tests presented here use ten realizations of the expected 5-year DESI data set.

The paper is organised as follows.
In Sect. \ref{sec:deltas} we show how continuum fitting results in a measured transmission field that is not equal to the true transmission field.
Section \ref{sec:ximeas} calculates the correlation functions of the measured
transmission field and describes how the physical model of the
true correlation function is corrected by using the distortion matrix.
Section \ref{sec:tests} describes tests of the distortion matrix
using mock data sets.
In section \ref{sec:fits}, the residual effects of the distortion are studied
and their effect on fit parameters is presented.
Section  \ref{sec:old} presents approximated distortion matrix that was first used in 
measurements of the correlation function
\citep{2017BautistaDR12,dmdB+17,2019deSainteAgatheDR14,2019BlomqvistDR14,dMdB+20,DESIyr1lyabao}
and how it compares with the matrix now used in DESI \citep{DESI-yr3lyabao}.
Section \ref{sec:concl} concludes.
In appendix \ref{sec:structure}, we give more details on the structure of the distortion matrix
and demonstrate the limitations of the current distortion matrix related to its extent in separation space.

\begin{figure}
\centering
\includegraphics[width=\textwidth]{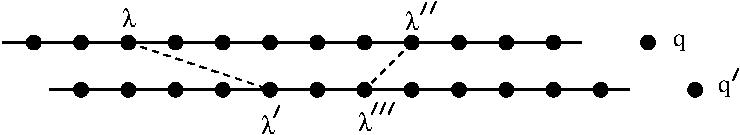}
\caption{
Two quasars and their forests.
  The measured correlation $\langle\dmql\dmqplp\rangle$ will have
  admixtures of all $\langle\delta^t_{q\lambda^{\prime\prime}}\delta^t_{q^\prime \lambda^{\prime\prime\prime}}\rangle$.
  If the intrinsic correlation $\langle\dtql\dtqplp\rangle$ is small,
  the measured correlations can be dominated by nearby pixels
  $(\lpp,\lppp)$ as illustrated here.
}
\label{fig:twoforests}
\end{figure}

\begin{figure}
\includegraphics[width=0.48\textwidth]{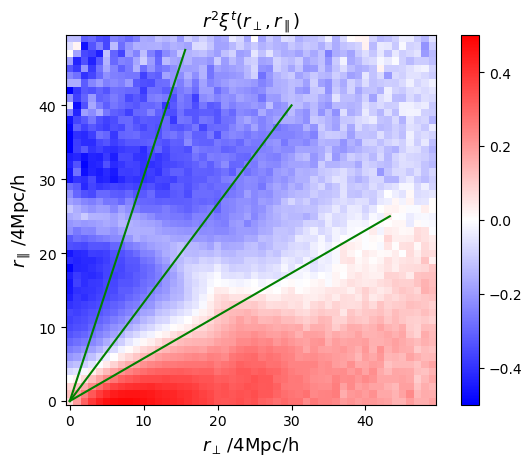}
\includegraphics[width=0.48\textwidth]{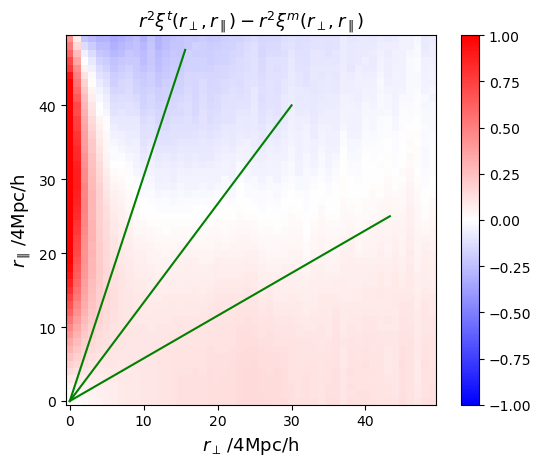}
\caption{Comparison of true and measured auto-correlation functions for the stack of 10 completed-survey DESI mock data sets \citep{Herrera_2024}.
The left panel shows the true auto-correlation function, $\xi^t(\rperp,\rpar)$
(multiplied by $r^2$ for better visualization),
and the right panel shows the difference between the true auto-correlation and
the measured auto-correlation, $\xi^m(\rperp,\rpar)$.
For small $\rperp$ and large $\rpar$, the large positive values of the difference  come about because
the continuum fitting effectively subtracts a portion of the large positive values of $\xi^t$ at small $r$.
For large $\rperp$, the positive (negative) differences at small (large) $\rpar$
result from subtracting a portion of  the neighboring positive (negative) correlations.
The green lines show the lines $\mu=\rpar/r=0.95$, 0.8 and 0.5.
}
\label{fig:xi_and_dxi}
\end{figure}

\section{The measured and true transmission fields}
\label{sec:deltas}

 In this section, we give formulae for the true transmission field (eqn. \ref{eq:deltadef}) and
the measured field (eqn. \ref{eq:deltahatdef}) found using continuum fitting.
We then use a simple flat-continuum model to illustrate the difference
between measured and true fields (eqn. \ref{eq:dhatexample}).
The "projected field" is then calculated by subtracting the mean and first moments of the field
for each quasar (eqn. \ref{eq:deltatildedef}) and we then show under what circumstances the correlations of the projected measured field are equal to the correlations of the projected true field.
We then show that this equality is realized in mock spectra.

The analysis of the \lya~forest uses 
the flux, $\fql$, of a quasar, $q$, at wavelength, $\lambda$.
which consists of a non-fluctuating component, $\Cql \Fzl$,
modulated by the fluctuating transmission field, $\dtql$:
\begin{equation}
    \fql=\Cql\; \Fzl\,(1+\dtql) 
\hspace*{5mm} \Rightarrow \hspace*{3mm} \dtql= \frac{\;\fql\;}{\Cql\,\Fzl} -1
    \label{eq:deltadef}
\end{equation}
Here, $\Cql$ is the unabsorbed "continuum", i.e. the spectrum in the absence of absorption, and $\Fzl$ mean absorption
at redshift $\zl=\lambda/\lambda_\alpha-1$.
The superscript of $\dtql$ refers to the "true" transmission field.

The eBOSS and DESI pipelines estimate 
the transmission field using the   
\texttt{picca}\footnote{https://github.com/igmhub/picca} \citep{picca_2021ascl.soft06018D} package which employs
a universal continuum template, $T(\lrf)$, a function of the restframe wavelength, $\lrf$. 
Each forest is  fit to the form ($a_q + b_q\Lql)\;\Tql$,
where $\Tql$ is the template redshifted to the quasar redshift and 
$\Lql=\log(\lambda)-\overline{\log(\lambda)}$
where  the overline refers to a (weighted) mean within a given forest, $q$.\footnote{In practice, \texttt{picca} uses $\Lql=\log(\lambda)-\log(\lambda)_{min}$.}
The parameters $(a_q,b_q)$ model the diversity and redshift evolution of quasar luminosity and spectral index.\footnote{The uncertainties in $(a_q,b_q)$ are not propagated to uncertainties in the correlation function which is determined globally by sub-sampling.}
An example of a DESI spectrum is shown in Fig. \ref{fig:oneforest} showing the measured flux as a function of wavelength, $\lambda$, and the estimated $\Cql\,\Fzl$ over the wavelength range used in this study: $1040<\lrf<1200\AA$.

The measured field is then
\begin{equation}
  \dmql= \frac{\fql + \nql}{(a_q + b_q\Lql)\;\Tql} -1
  \hspace*{10mm}
  \label{eq:deltahatdef}
\end{equation}
where we have added a noise term $\nql$.

As a illustrative example, we suppose that
the spectrum is flat  apart from fluctuations:
$\fql=f_0(1+\dtql)$.
Using a flat template, $\Tql=1$,
a least-squares fit of  a spectrum with $N$ points of weight $\wql=1$
gives
\begin{equation}
a_q=f_0[1+ N^{-1}\sum_\lambda (\dtql +\nql/f_0)] \hspace*{5mm} 
b_q=f_0 \frac{\sum_\lambda \Lql (\dtql+\nql/f_0)}{\sum_\lambda\Lql^2} \hspace*{5mm}
\end{equation}
In the absence of fluctuations and noise, $a_q=f_0$ and $b_q=0$.
We inject the expressions for $a_q$ and $b_q$ into  (\ref{eq:deltahatdef}) giving
\begin{equation}
    \dmql=\frac{1+\dtql+\nql/f_0}{1+\eql} -1
\end{equation}
where
\begin{equation}
\eql=N^{-1}\sum_{\lambda^\prime}(\dtqlp+\nqlp/f_0)(1+\Lql\Lqlp/\overline{\Lambda^2}) 
\end{equation}
is the mean fluctuation in the forest weighted by the $\Lql\Lqlp$-dependent factor. 
DESI uses  pixels of width $\Delta\lambda=0.8\AA$ giving $N\approx600$.
The noise of the completed DESI survey results in a r.m.s. of
$\eql$ of order 0.05.
This allows as to  expand the denominator, giving 
\begin{equation}
   \dmql = \dtql + \nql/f_0 -\eql -\eql(\dtql +\nql/f_0) 
   \;\;+O(\eql^2)
   \label{eq:dhatexample}
\end{equation}
We have $\dtql=\dmql$ only in the absence of noise and in the limit $\eql\rightarrow0$.
The third term on the right-hand side, $\eql$, shows how the $\dmql$ are
a mixture of  the $N$ $\dtqlp$ in the forest.
Of some importance that we will see below, the difference $\dmql-\dtql$  due to this term is linear in $\Lql$.
The term $\eql\dtql$ is the mean correlation between pixels of the same forest and is of order 1\% for DESI forests. The resulting three-point terms in the measured inter-forest correlation are therefore expected to be small.

While we have derived eqn. \ref{eq:dhatexample} with a very simple model, it is qualitatively useful in the general case. In practice, pixel weights, $\wql$, are used so the mean fluctuation appropriate for $\eql$ is the weighted mean fluctuation.
The use of a known wavelength-dependent template does not change eqn. \ref{eq:dhatexample}.  However, the assumption of the uniformity of the template is not realistic because of quasar spectral diversity.
An important assumption of the distortion matrix program is therefore that spectral diversity is due to local effects and does not lead to long-range correlations.

 The $\eql$ terms in equation \ref{eq:dhatexample} show that the measured field, $\dmql$ is a mixture of all the true-field elements, $\dtqlp$, in the forest.
Therefore, for two forests with $N_q$ and $N_{q^\prime}$ wavelengths,
each ``distorted'' correlation $\langle\dmql\dmqplp\rangle$
has $N_qN_{q^\prime}$ terms
linear in the true correlation
$\langle\delta^t_{q\lambda^{\prime\prime}}\delta^t_{q^\prime\lambda^{\prime\prime\prime}}\rangle$
plus 3-point terms and terms of $O(N^{-2})$ and higher.
As illustrated in Fig. \ref{fig:twoforests},
if the intrinsic correlation $\langle\dtql\dtqplp\rangle$ is small
because of a large $\lambda-\lambda^\prime$,
the measured correlations can be dominated by nearby pixels
$(\lpp,\lppp)$, leading to significant distortion.

The auto-correlation function is generally calculated by regrouping $(q\lambda,q^\prime\lambda^\prime)$ pairs in bins of $(\rperp,\rpar)$, 
where $\rperp$ and $\rpar$ are the separations perpendicular to and along the line of sight. 
Figure \ref{fig:xi_and_dxi} shows, for the stack of 10 mocks, the true auto-correlation function (on the left) and the difference between the true and measured correlations (on the right).  As expected from Fig. \ref{fig:twoforests}, there are large differences    at small $\rperp$. 
At small $\rperp$ and large $\rpar$, we have $\xi^t>\xi^m$ because the mixing subtracts a fraction of the large positive true correlations at small $\rpar$.
At large $\rperp$ this distortion comes from subtracting the mean of neighboring
correlations which, as seen in the left panel, are positive at small $\rpar$ and negative at large $\rpar$.

The \texttt{picca} procedure for determining the $(a_q,b_q)$ approximately drives
the mean and first moments of the $\dmql$ to zero.  However, instead of
using the thus calculated $\dmql$, the \texttt{picca} procedure
uses the "projected" measured field $\dmpql$:
\begin{equation}
\dmpql= \sum_{\lp} \eta_{q\lambda\lambda^\prime}\dmqlp
\hspace*{10mm}
\eta_{q\lambda\lambda^\prime}=\delta^K_{\lambda\lambda^\prime}
\,-\, 
  \frac{\wqlp}{\Wq} 
\left[ 1 + \frac{\Lql\Lqlp}{\overline{\Lql^2}}\right]
\label{eq:deltatildedef}
\end{equation}
where  $\delta^K$ is the Kronecker delta and  $\Wq=\sum\wql$ 
is the sum of the pixel weights.
The mean and first moment of the $\dmpql$ are exactly zero.
This is done because it is possible to calculate the correlation function
of the projected true field, $\dtpql$, and it is reasonable to assume
that the correlations of the $\dtpql$ are nearly equal to those of the $\dmpql$.
This is true if the measured
field, $\dmql$, differs from the true field by a  linear function 
of $\Lql$ (as is the case of eqn. \ref{eq:dhatexample}):  
\begin{equation}
\dmql-\dtql=A_{q0} + A_{q1}\Lql \;.
\label{eq:condition}
\end{equation}
In the calculation of the projected field $\dmpql$, the terms with $A_{q0}$ and $A_{q1}$
cancel leaving
\begin{equation}
  \dmpql=
  \sum_{\lp} \eta_{q\lambda\lambda^\prime}\dtqlp \;=\; \dtpql
  \label{eq:idealcase}
\end{equation}
We see that in this case, the projection of the measured field, $\dmpql$, 
is equal to the projection of the true field, $\dtpql$.
It follows that the correlations of the projected measured field are
equal to the correlations of the projected true field.

We cannot expect condition~\ref{eq:condition} to be perfectly respected
for each forest, first, because of noise and, second, because
the peculiarities of individual
quasar spectra means that the template is imperfect
and the imperfections will contribute to $\dmql-\dtql$.
We expect that the noise is uncorrelated between 
different forests and can be ignored.
Furthermore, if the template imperfections 
are caused by local astrophysical 
effects on quasar spectra, they will not contribute to correlations
between forests or between forests and quasars. 

The basic assumption of our treatment of continuum-fitting distortion is then
the  
near equality of the projected-true and projected-measured correlations, i.e. that the template imperfections and noise are uncorrelated and that the higher-order terms in \ref{eq:dhatexample} can be neglected.
This hypothesis can be tested using  the mock data sets where both the measured and
true fields are known.

\begin{figure}
    \centering    
    \includegraphics[width=0.43\textwidth]{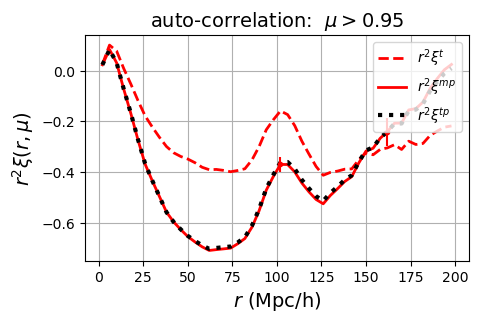} 
    \includegraphics[width=0.43\textwidth]{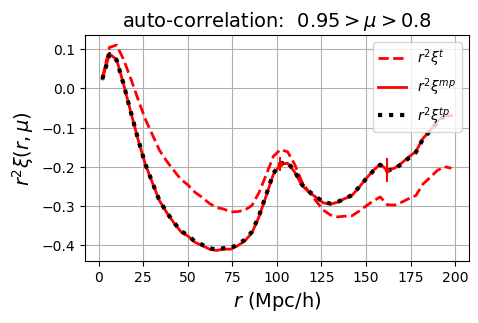}\\
    \includegraphics[width=0.43\textwidth]{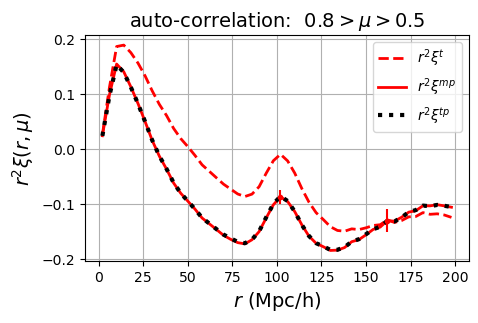}
    \includegraphics[width=0.43\textwidth]{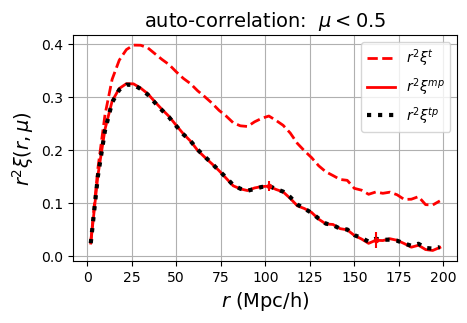} \\
    \includegraphics[width=0.43\textwidth]{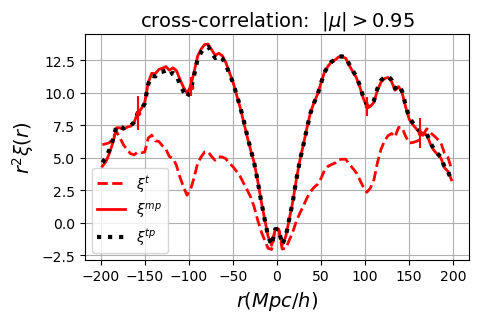}
    \includegraphics[width=0.43\textwidth]{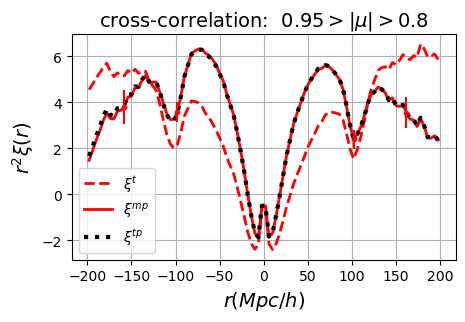}\\
    \includegraphics[width=0.43\textwidth]{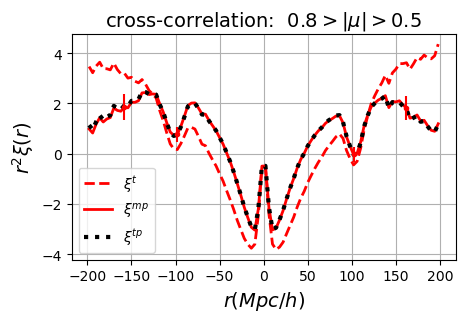}
    \includegraphics[width=0.43\textwidth]{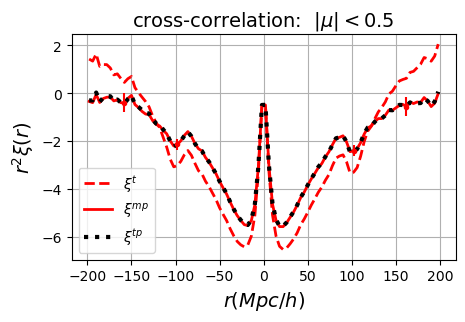}
    \caption{The auto-correlation (top four panels) and the cross-correlation (bottom four panels) for the stack of 10 completed-survey DESI mock data sets \citep{Herrera_2024}. The solid lines show the correlations of the  projected measured field, $\dmpql$, averaged over the $\mu$ range as labeled.  The black dotted lines show the measured correlation of the projected true field, $\dtpql$.
    The dashed lines show the correlations of the true field, $\dtql$.  
    The error bars are
    representative of the expected DESI results.
    For the cross-correlation, points with $\mu<0$ are plotted at $r<0$.}
    \label{fig:xi_projected_true}
\end{figure}

The results are sumarized in Fig. \ref{fig:xi_projected_true} which shows the ten-mock stack of the auto- and cross-correlation in four 
ranges of $\mu=\rpar/r$.
The dashed lines are the true correlations 
and the corresponding
solid lines are the correlations of the projected measured field.  The black-dotted lines show the
correlations of the projected true field.  The agreement between the solid and dotted lines
indicates that our assumption of the near equality of the projected-measured and projected-true correlations is reasonable for these mocks.
Application of this assumption to real data supposes that correlated noise and
astrophysical effects are small.

\section{The measured correlation function and the distortion matrix}
\label{sec:ximeas}

The correlation functions of the projected measured field are measured
by using a fiducial cosmological model to transform redshift separations
and angular separations into distance separations, $(\rperp,\rpar)$,
perpendicular to and parallel to the line of sight.
The estimators of the correlation functions in a $(\rperp,\rpar)$ bin $A$ are:
\begin{equation}
	\xi_{A}^{auto} = \frac{1}{W^{auto}_A}
    \sum\limits_{(q\lambda,q^\prime\lambda^\prime) \in A} \wql \wqplp \, \dmpql \dmpqplp
    \hspace*{10mm}
W^{auto}_A=	\sum\limits_{(q\lambda,q^\prime\lambda^\prime) \in A} \wql \wqplp
    \label{eq:autoestimator}
    \end{equation}
    \begin{equation}
    \xi_{A}^{cross} = \frac{1}{W^{cross}_A} 
    \sum\limits_{(q\lambda,Q) \in A} \wql w_Q \, \dmpql
    \hspace*{10mm}
W^{cross}_A=	\sum\limits_{(q\lambda,Q) \in A} \wql w_Q .
    \label{eq::crossestimator}
\end{equation}
For the auto-correlation,
the sum is over pixel pairs in the $(\rperp,\rpar)$ bin $A$ and
the $\wql$ are pixel weights chosen to optimize the measurement.
For the cross correlation, the sum is over quasar
pixel pairs with $w_Q$ the weight for the quasar.
BOSS-eBOSS and DESI have used  $(\rperp,\rpar)$ bins of size
$4\,{\rm Mpc/h}\times 4\,{\rm Mpc/h}$.

We want to determine which physical model best fits the measured auto- and cross-correlation functions.
From eqn. \ref{eq:idealcase}, the correlations of the best model after projection should approach
the measured correlations.
A brute force method would then be to generate  a large number of mock data sets according to a variety of  models, 
project the model $\dtql$'s, calculate the correlation functions of the $\dtpql$'s, and
then compare with the measured correlations.
Fortunately, there is a way of avoiding this time-consuming procedure by directly calculating
the correlations of the projected model.
Using eqn. \ref{eq:idealcase} we have
\begin{equation}
	\xi_{A}^{auto} = W_A^{-1}
    \sum\limits_{(q\lambda,q^\prime\lambda^\prime) \in A} \wql \wqplp \, 
    \sum_{\lambda^{\prime\prime}}\eta_{q\lambda\lambda^{\prime\prime}}\delta^t_{q\lambda^{\prime\prime}}
    \sum_{\lambda^{\prime\prime\prime}}\eta_{q^\prime\lambda^\prime\lambda^{\prime\prime\prime}}\delta^t_{q^\prime\lambda^{\prime\prime\prime}}
    \label{eq:autoestimator2}
\end{equation}
The product $\delta^t_{q\lambda^{\prime\prime}}\delta^t_{q^\prime\lambda^{\prime\prime\prime}}$
can be replaced with its expectation value, 
$\xi^{at}(q\lambda^{\prime\prime},q^\prime\lambda^{\prime\prime\prime})$.
\begin{equation}
 \delta^t_{q\lambda^{\prime\prime}}\delta^t_{q^\prime\lambda^{\prime\prime\prime}} \;\rightarrow\;
 \xi^{at}(q\lambda^{\prime\prime},q^\prime\lambda^{\prime\prime\prime}) \;=\;
 \xi^{at}_{A^\prime}(z_0) 
 f^{auto}_{A^\prime}(\lambda^{\prime\prime},\lambda^{\prime\prime\prime},z_0)
 \label{eq:replacement}
\end{equation}
 where for $\xi^{at}_{A^\prime}$ the superscript is short for "auto-true",
$A^\prime$ is the $(\rperp,\rpar)$ bin corresponding to the pair 
$(q\lambda^{\prime\prime},q^\prime\lambda^{\prime\prime\prime})$, $z_0$ is a reference redshift and
$f^{auto}_{A^\prime}(\lambda^{\prime\prime},\lambda^{\prime\prime\prime},z_0)$ accounts for the redshift evolution of the correlation function between the reference redshift and redshift of the measurement.
Regrouping terms to replace the sums over $\lambda^{\prime\prime}$ and $\lambda^{\prime\prime\prime}$
with a sum over $A^\prime$ and
$(q\lambda^{\prime\prime},q^\prime\lambda^{\prime\prime\prime})\in A^\prime$ we obtain
\begin{equation}
	\xi_{A}^{auto} =  \sum_{A^\prime} D^{auto}_{AA^\prime} \; \xi^{at}_{A^\prime}(z_0)
\end{equation}
where the distortion matrix is defined by 
\begin{equation}
    D^{auto}_{AA^\prime} \;=\;
    W_A^{-1}
    \sum\limits_{(q\lambda,q^\prime\lambda^\prime) \in A} \wql \wqplp \, 
    \sum_{(q\lambda^{\prime\prime},q^\prime\lambda^{\prime\prime\prime})\in A^\prime}\eta_{q\lambda\lambda^{\prime\prime}}
    \eta_{q^\prime\lambda^\prime\lambda^{\prime\prime\prime}}        
f^{auto}_{A^\prime}(\lambda^{\prime\prime},\lambda^{\prime\prime\prime},z_0)
    \label{eq:dmauto}
\end{equation}
The parameters of the mock generator \citep{Farr+19,colore_2022JCAP...05..002R,Herrera_2024} were adjusted so that
the redshift evolution of the auto-correlation function was 
\begin{equation}
f^{auto}_{A^\prime}(\lambda^{\prime\prime},\lambda^{\prime\prime\prime},z_0) = 
\left( \frac{1+z^{\prime\prime}}{1+z_0}  \right)^{\gamma_\alpha -1} 
\left( \frac{1+z^{\prime\prime\prime}}{1+z_0}  \right)^{\gamma_\alpha -1} 
\hspace*{10mm}
\gamma_\alpha = 2.9
\label{eq:auto_zevolution}
\end{equation}
where $(z^{\prime\prime},z^{\prime\prime\prime})$ are the redshifts of 
$(\lambda^{\prime\prime},\lambda^{\prime\prime\prime})$ assuming \lya~absorption.

For the cross-correlation, one finds
\begin{equation}
    D^{cross}_{AA^\prime} \;=\;
    W_A^{-1}
    \sum\limits_{(q\lambda,Q) \in A} \wql w_Q \, 
    \sum_{(q\lambda^{\prime},Q)\in A^\prime}\eta_{q\lambda\lambda^{\prime}}
    f^{cross}_{A^\prime}(\lambda^{\prime},z_Q,z_0)
    \label{eq:dmcross}
\end{equation}
\begin{equation}
f^{cross}_{A^\prime}(\lambda^\prime,z_Q,z_0) = 
\left( \frac{1+z^\prime}{1+z_0}  \right)^{(\gamma_\alpha -1)} 
\left( \frac{1+z_Q}{1+z_0}  \right)^{(\gamma_Q -1)} 
\hspace*{10mm}
\gamma_Q = 1.44
\label{eq:cross_zevolution}
\end{equation}
where $z^\prime$ is the redshift of $\lambda^\prime$.

The distortion matrices $D^{auto}_{AA^\prime}$ and $D^{cross}_{AA^\prime}$ have
an index $A$ referring to a bin of the measured correlation function and an 
index $A^\prime$ referring to a bin of the true correlation function.
For the analysis presented here, the bins of $A$ are of width $4\hMpc$ in both directions covering the range 
$0<\rperp<200\hMpc$ and $0<\rpar<200\hMpc$ (auto-corrrelation) and
$0<\rperp<200\hMpc$ and $-200<\rpar<200\hMpc$ (cross-correlation).
(Negative $\rpar$ is for quasars nearer then the forest pixel.)
The $4\hMpc$ wide bins of $A^\prime$ for this study cover an extended $\rpar$ range:
$0<\rpar<300\hMpc$ (auto-corrrelation) and
$-300<\rpar<300\hMpc$ (cross-correlation).
The extended range for $A^\prime$ is necessary because typical DESI forests have a total length of $400\hMpc$ resulting in measured correlations that are mixed with correlations
of much greater distance.
The BOSS-eBOSS studies used maximum values of $|\rpar|$ of $200\hMpc$ for $A^\prime$ and subsequent studies showed that the distortions had not converged for $\rpar>150\hMpc$.
This motivated the decision in DESI to extend the matrix to $\rpar=300\hMpc$.

The elements of the two distortion matrices can be found
by averaging over a representative sample of quasar-pixel
or pixel-pixel pairs; 
about 1\% of the pairs is sufficient.
The diagonal elements are very near unity, corresponding to the mixing of a given
pixel with itself.
Because neighboring forests are nearly parallel, the off-diagonal elements vanish
unless 
$r_{\perp A}\approx r_{\perp A^\prime}$ 
in which case they
are of order a few percent.
Examples are shown in Appendix \ref{sec:structure} in Figs.~\ref{fig:xdm_cf} and \ref{fig:dm_cf}.
As illustrated by fig. \ref{fig:twoforests}, the small size of the off-diagonal 
terms does not mean that the distortion is small because they may multiply
a correlation that is much larger than the diagonal correlation.

\begin{figure}
    \centering
    \includegraphics[width=0.45\linewidth]{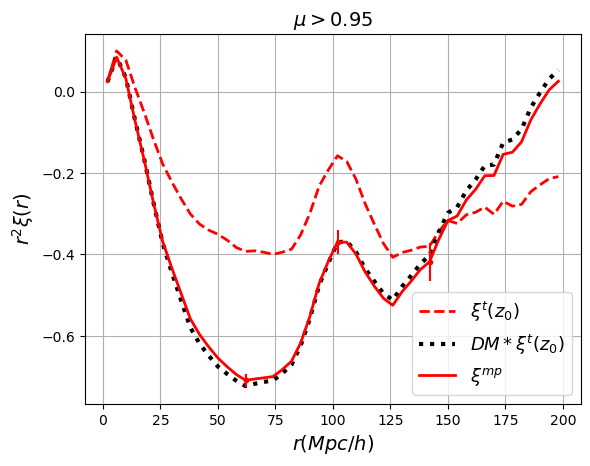}
    \includegraphics[width=0.45\linewidth]{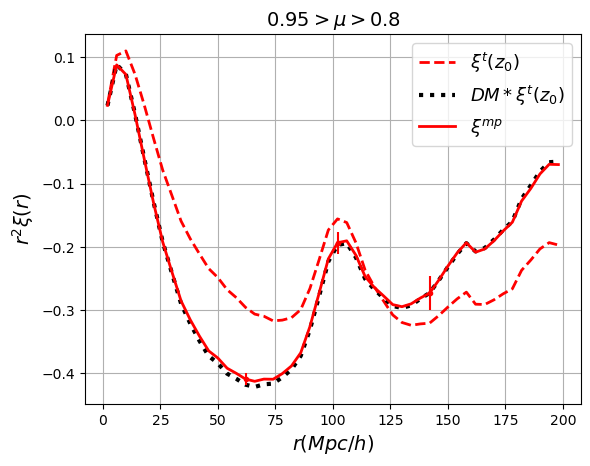}\\
    \includegraphics[width=0.45\linewidth]{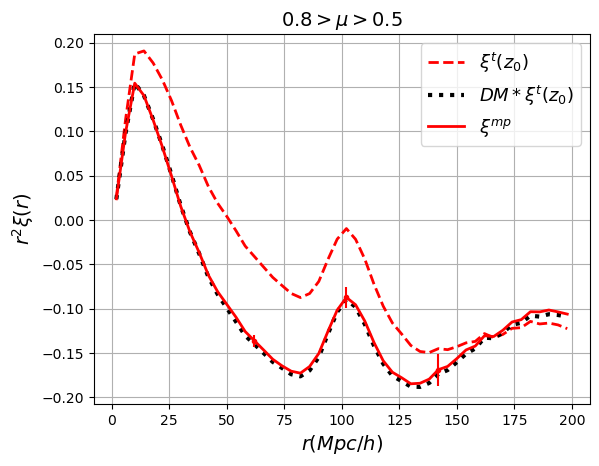}
    \includegraphics[width=0.45\linewidth]{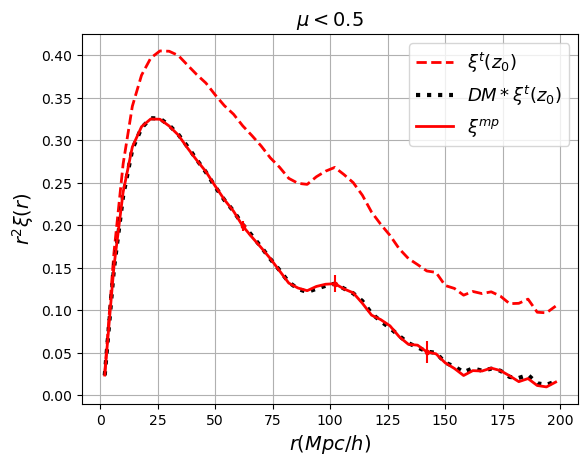}\\
    \caption{
    The auto-correlation function for the stack of 10 completed-survey DESI mock data sets. The solid  red lines show the correlations of the  projected measured field, $\dmpql$, averaged over the $\mu$ range as labeled.  The dashed red lines show the measured correlation of the true field, $\dtql$.
    The dotted black  lines show the correlations of the true field corrected by the distortion matrix, $DM*\xi^{t}(z_0)$.  
    The error bars are
    representative of the expected completed-survey DESI results.}
    \label{fig:zcfwedges}
\end{figure}

\begin{figure}
    \centering
    \includegraphics[width=0.45\linewidth]{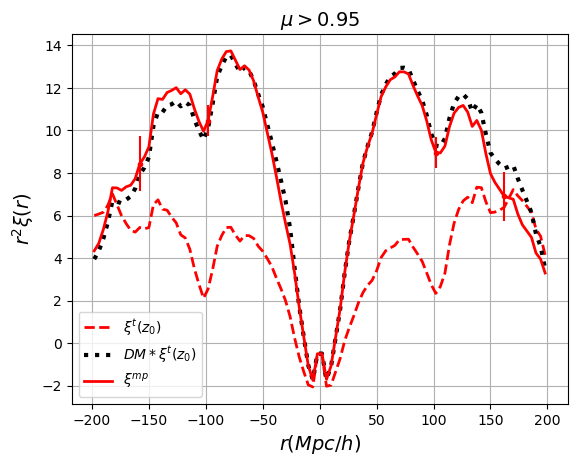}
    \includegraphics[width=0.45\linewidth]{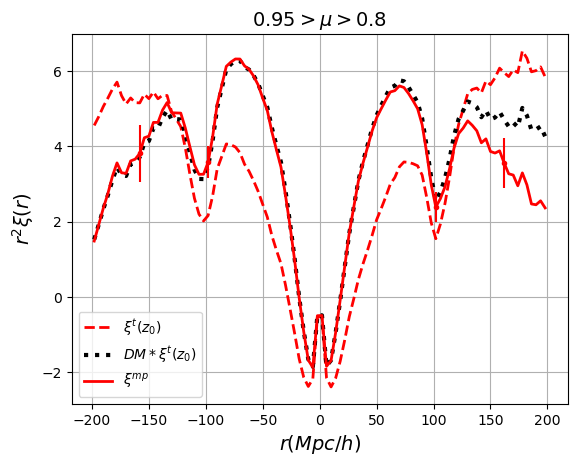}\\
    \includegraphics[width=0.45\linewidth]{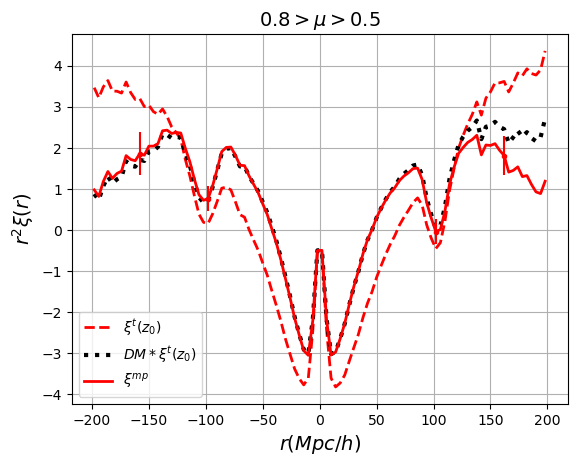}
    \includegraphics[width=0.45\linewidth]{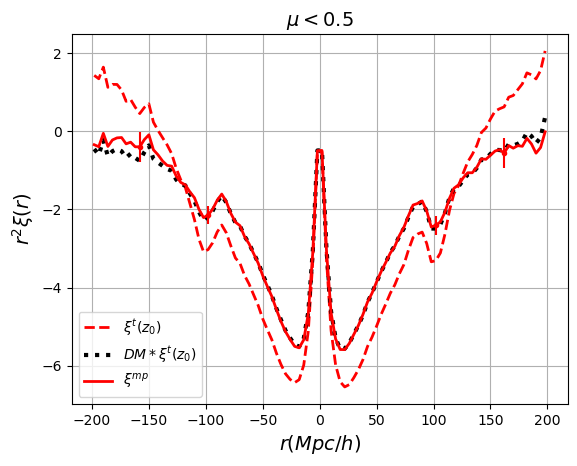}\\
    \caption{ The same as Fig. \ref{fig:zcfwedges} except now for 
the cross-correlation function for the stack of 10 completed-survey DESI mock data sets. 
Points with $\mu<0$ are plotted at $r<0$.
    }
    \label{fig:zcxfwedges}
\end{figure}

\begin{figure}
    \centering
    \includegraphics[width=0.45\linewidth]{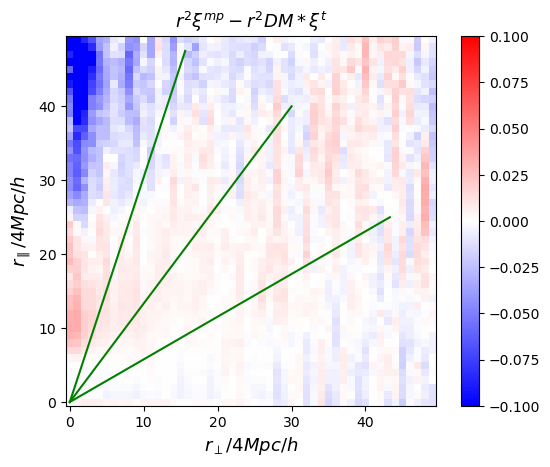}
    \includegraphics[width=0.45\linewidth]{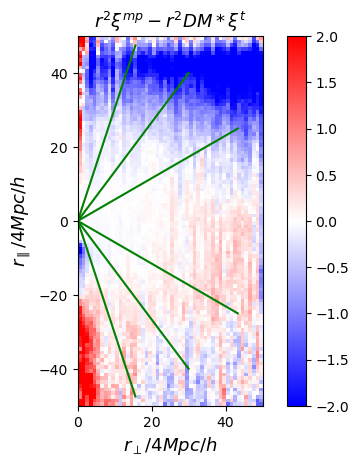}\\
    \caption{ 
    The difference between the measured projected correlation function, $r^2\xi^{mp}$
    and its value predicted by the distortion matrix, $r^2DM*\xi^{t}$,
    for the auto-correlation (right) and cross-correlation (left).
    The green lines show the lines $|\mu|=|\rpar|/r=0.95$, 0.8 and 0.5.
    }
    \label{fig:drsxi_vs_rtrp}
\end{figure}

\begin{figure}
    \centering
    \includegraphics[width=0.45\linewidth]{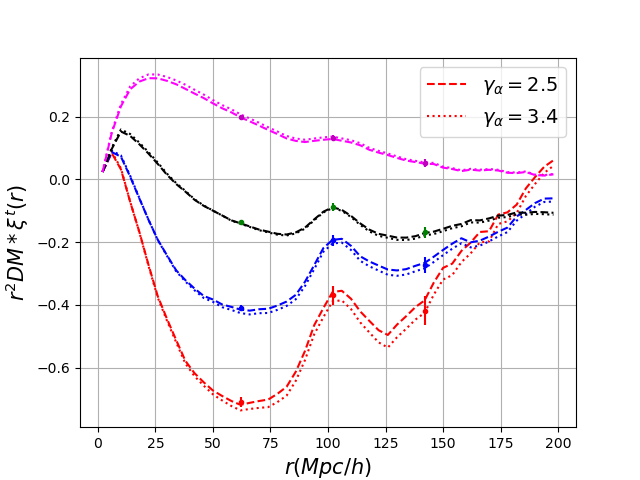}
    \includegraphics[width=0.45\linewidth]{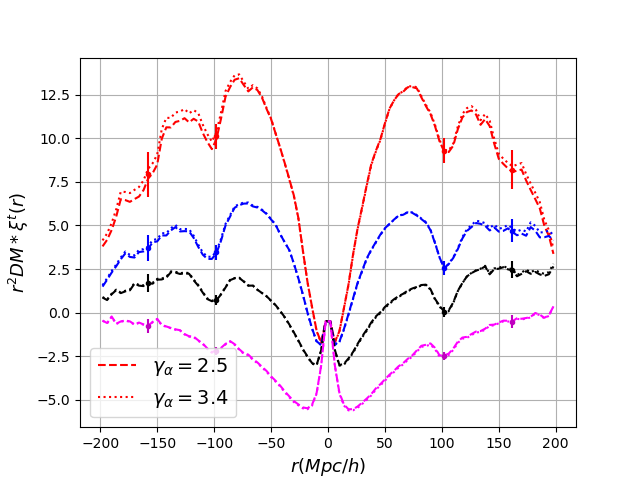}\\
    \caption{ The effect of the assumed $\gamma_\alpha$ on the calculated correlations. 
The auto- and cross-correlation functions are plotted in four ranges
of $\mu$:
$|\mu|>0.95$ (red), $0.8<|\mu|<0.95$ (blue), $0.5<|\mu|<0.8$ (black), and $|\mu|<0.5$ (magenta).
The solid and dashed lines show the true correlations corrected by the distortion matrices 
for the two non-standard values of $\gamma_\alpha$ as labeled}
    \label{fig:gamma}
\end{figure}

\section{Tests of the distortion matrix}
\label{sec:tests}

We tested the distortion matrix paradigm using  mock
spectra where the true unabsorbed continuum, $\Cql$, 
and mean transmission, $\Fzl$, 
are known, allowing
the true $\dtql$ to be calculated using eqn. \ref{eq:deltadef}.
The auto- and cross-correlation functions of the $\dtql$
can then be multiplied by the distortion matrix and
directly compared with the correlations of the $\dmpql$.

We used \texttt{LyaCoLoRe} mocks as described in \citep{Farr+19,colore_2022JCAP...05..002R,Herrera_2024}.
They were produced using Gaussian-random fields to which the Fluctuating Gunn-Peterson approximation was used to generate \lya~absorption in the intergalactic medium.
The footprint and spectral noise were chosen so as to simulated the completed-survey
DESI data.
Ten independent realizations were used.
For these mocks, absorption by metals or by high-column density systems was not included.
Continuum templates were generated using the \texttt{simqso} library\footnote{https://github.com/desihub/simqso} \citep{simqso_2021ascl.soft06008M}.
The model produces continua with a broken power-law shape on which are superimposed a set of Gaussian emission
lines defined by their wavelength, equivalent width, and Gaussian RMS width.
As described in \citep{Herrera_2024}, the parameters of these two components are randomly chosen so as to reflect the diversity of
DESI quasars.

In calculating the correlations of the $\dtql$, we must confront 
an unfortunate feature  of the \lya~forest analysis: the $\dtql$ do not randomly sample the IGM because they must be placed in front of a quasar with which they are correlated.  This means that the mean of the 
$\dtql$ does not vanish: $\overline{\dtql}\approx7\times10^{-4}$ and this must be taken
into account by subtracting this value before calculating the correlations.
This does not affect the correlations of the $\dmpql$ and $\dtpql$ because the projection
effectively subtracts the mean forest by forest.
For the correlations of $\dtql$, it is necessary 
to first subtract 
$\overline{\dtql}$ before calculating the
correlations and multiplying by the distortion matrix.
The small size of $\overline{\dtql}$  means that this has no measurable effect 
on the auto-correlation but the subtraction is necessary to insure that the cross-correlation vanishes
at large distances.  

The test of the distortion matrix then consists of multiplying
the auto- and cross-correlation functions of the $\dtql$
by the distortion matrix and
directly comparing them with the correlations of the $\dmpql$.
This is done in Figs. \ref{fig:zcfwedges} and \ref{fig:zcxfwedges}.
Comparing the solid-red and dotted-black lines,
one sees that the distortion matrix captures the majority 
of the effect, with the exception of large positive $\rpar$ for the
cross-correlation.  
This is seen in $(\rperp,\rpar)$ space in Fig. \ref{fig:drsxi_vs_rtrp}.
The large imperfections seen at low $\rperp$ are due to the fact that the corrections are large for low $\rperp$.
The large imperfections at large $\rpar$ for the cross-corelation may be due to
truncating the distortion matrix at $\rpar=300\hMpc$ and could be perhaps eliminated
by increasing the size of the distortion matrix in the $\rpar$ direction.
A detailed discussion of this problem is given in appendix \ref{sec:structure}.

The calculations of the distortion matrices, equations \ref{eq:dmauto} and \ref{eq:dmcross},
have two sources of error. The first is the statistical uncertainty coming from sampling the pairs and can be made negligible by using a sufficient number of pairs.
The second source is the uncertain knowledge of the redshift evolution of the correlations, parameterized by $\gamma_\alpha$ and $\gamma_Q$ in equations \ref{eq:auto_zevolution} and \ref{eq:cross_zevolution}.
Figure \ref{fig:gamma} shows the effect of varying $\gamma_\alpha$ over the range $[2.5,3.4]$.
While measureable with mocks, it is a small effect.
Its effect on the data will require a determination of the uncertainty of $\gamma_\alpha$ and $\gamma_Q$

\section{Effect of distortion on fit parameters}
\label{sec:fits}

Figure \ref{fig:drsxi_vs_rtrp} 
shows that the distortion-matrix correction to the correlation functions is not perfect.
It is therefore important to see if these imperfections have a significant
impact on the model best-fit parameters.
We investigated this
by comparing the standard fits using the distortion matrix  with
fits using the true-continuum correlations with the distortion matrix set
to unity. 
The physical parameters of the model are the forest flux bias, $\blya$, the redshift-space distortion parameter, $\betalya$,
and the two BAO-peak position parameters $(\aperp,\apar)$ that describe the relative
shift of the peak in the $(\rperp,\rpar)$ direction from the peak position of the fiducial model.
For the fits of the cross-correlation, the quasar bias was fixed to the value $b_Q=4.04$ since
only the product of the quasar and forest biases is determined by the fit.

Table \ref{tab:bestfit} shows the fit values for the uncorrelated combinations of  
$(\blya,\betalya)$ and of $(\aperp,\apar)$ fit over the range $10<r<180\hMpc$.
The effective bias, $\blya(1+0.7\betalya)\approx0.275$, is stable at the sub-percent level
for the true-continuum and standard fits, and for both the auto- and cross-correlations.
On the other hand, $\betalya$ is $\approx4\%$ lower for the standard fits than for the 
true-continuum fits, indicating that systematic effects due to the continuum-fitting distortion may have a small effect on the measurement of $\betalya$.

The uncorrelated BAO parameters $0.6\apar+0.4\aperp$  and $0.4(\apar-1)-0.6(\aperp-1)$
are all consistent with the expected values $(1,0)$ within two standard deviations.
The most significant deviation is for $0.4(\apar-1)-0.6(\aperp-1)$ with is $\approx1.7$
standard deviations from zero for the standard fits on both the auto- and cross-correlations.
We note that the extensive mock studies for the DESI DR1 and DR2 BAO measurements \citep{DESIyr1lyabao,DESI-yr3lyabao} have
shown no biases at a much more precise level, so the small effect seen here is likely a 
statistical fluctuation due to the small number of mocks used.

\begin{table}
\begin{center}
\begin{tabular}{|c|c|c|}
\hline
& true continuum & standard \\
\hline
 $\blya(1+0.7\betalya)$ (auto) & 0.2774 $\pm$ 0.0001 & 0.2762 $\pm$ 0.0001\\ 
 $\blya(1+0.7\betalya)$ (cross) & 0.2733 $\pm$ 0.0003 & 0.2728 $\pm$ 0.0003\\ 
 $\betalya$ (auto)&  $ 1.528  \pm  0.004 $ &  $1.452  \pm  0.005 $ \\
 $\betalya$ (cross) &  $ 1.519  \pm  0.012 $  &  $1.463  \pm  0.008 $ \\ 
 $0.6\apar+0.4\aperp$ (auto)& 0.9979 $\pm$ 0.0015 & 0.9998$\pm$  0.0013 \\ 
 $0.6\apar+0.4\aperp$ (cross)& 0.9993 $\pm$ 0.0011 & 1.0005$\pm$  0.0012 \\ 
 $0.4(\apar-1)-0.6(\aperp-1)$ (auto)& 0.0003$\pm$  0.0024 & 0.0060 $\pm$ 0.0036 \\ 
 $0.4(\apar-1)-0.6(\aperp-1)$ (cross)& 0.0031$\pm$   0.0024 & 0.0052 $\pm$ 0.0032 \\ 
\hline 
\end{tabular}
\vspace*{2mm}
\caption{
Best fit values of the uncorrelated combinations of $(\blya,\betalya,\aperp,\apar)$ for the true-continuum
mocks with unit distortion matrix and the  standard analysis.
The values given are the mean of 10 mock realization of 5 years of DESI
and the uncertainties are the dispersions divided by $\sqrt{10}$.
}
\label{tab:bestfit}
\end{center}
\end{table}

\section{The original distortion matrix}
\label{sec:old}

Prior to the DR2 DESI analysis \citep{DESI-yr3lyabao},
the  BOSS-eBOSS \citep{2017BautistaDR12,dmdB+17,2019deSainteAgatheDR14,2019BlomqvistDR14,dMdB+20} and the DESI DR1 \citep{DESIyr1lyabao} analyses used a simplified distortion matrix.
Instead of the replacement given by eqn. \ref{eq:replacement},
these analyses  used
\begin{equation}
 \delta^t_{q\lambda^{\prime\prime}}\delta^t_{q^\prime\lambda^{\prime\prime\prime}} \;\rightarrow\;
 \xi^{at}_{A^\prime}
 \label{eq:replacement_old}
\end{equation}
This replacement makes the assumption that the pairs
$(q\lambda^{\prime\prime},q^\prime\lambda^{\prime\prime\prime})$
used in the determination of $\xi_A$ are representative of the
pairs used in the determination of $\xi_{A^\prime}$.
This leads to correlations functions given by
\begin{equation}
\xi^{auto}_A = \sum_{A^\prime}D^{auto}_{AA^\prime}\xi^{at}_{A^\prime}
\hspace*{10mm}
D_{AA^\prime}^{auto}=
W_A^{-1}\sum_{(q\lambda,q^\prime\lambda^\prime)\in A} \wql \wqplp 
\sum_{(q\lpp,q^\prime\lppp) \in A^\prime}
\eta_{q\lambda,q\lpp}
\eta_{q^\prime\lp,q^\prime\lppp}
\label{eq:oldauto}
\end{equation}

\begin{equation}
\tilde\xi^{cross}_A = 
\sum_{A^\prime}D^{cross}_{AA^\prime}\xi^{ct}_{A^\prime}
\hspace*{10mm}
D_{AA^\prime}^{cross} = W_A^{-1}\sum_{q\lambda,Q \in A} 
w_{Q\lambda}  \sum_{q\lambda^\prime,Q \in A^\prime}
\eta_{q\lambda\lambda^\prime}
\label{eq:oldcross}
\end{equation}

\begin{figure}
\includegraphics[width=0.48\textwidth]{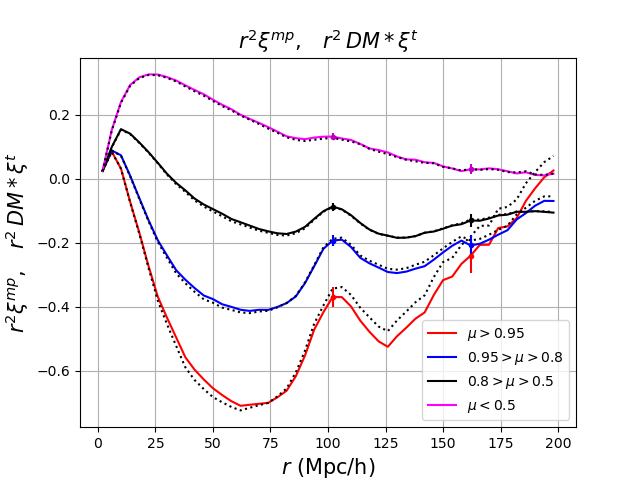}
\includegraphics[width=0.48\textwidth]{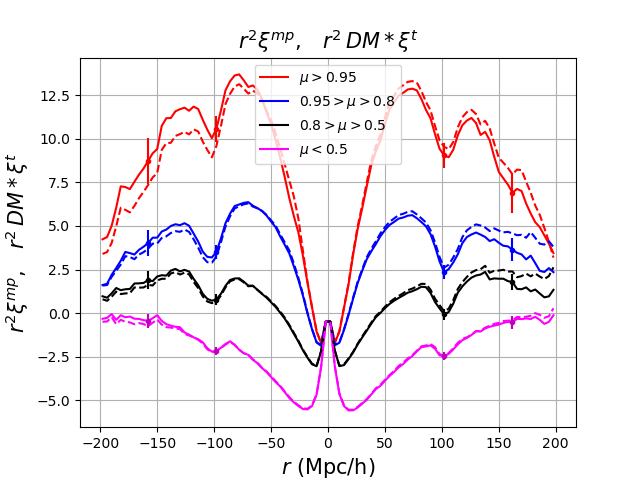}
\caption{Test of the original distortion matrices, to be compared with the test
of the present distortion matrix in Figs. \ref{fig:zcfwedges} and \ref{fig:zcxfwedges}.
The auto- and cross-correlation functions are plotted in four ranges
of $\mu$ as labeled.
The solid lines show the correlations of the measured projected field, $\dmpql$ and
the dotted black lines the true correlations corrected by the
original distortion matrices defined by eqns. \ref{eq:oldauto} and \ref{eq:oldcross}.
}
\label{fig:cfwedges} 
\end{figure}

The test of the old distortion matrix is shown  in Fig.~\ref{fig:cfwedges}.
Comparing with Fig.~\ref{fig:zcfwedges}, we see that the new matrix performs
better than the original for $\mu>0.95$.
For the cross-correlation, the problem at large positive $\rpar$ is seen
in both the new and original matrices.
In the DESI DR2 analysis \citep{DESI-yr3lyabao}, it was shown that the BAO-peak parameters derived using the original and new distortion matrices had no measurable difference.

\section{Conclusions}
\label{sec:concl}

The tests with mock spectra presented here show that the distortion matrix formalism
captures most of the distortion caused by continuum fitting, especially at distances
near or below the BAO peak.
The residual discrepancies, seen in Fig. \ref{fig:drsxi_vs_rtrp}
may lead to small shifts in the bias parameters $(\blya,\betalya)$ but
are unlikely to significantly affect the measurement of the BAO peak position at more than the percent level.
An obvious improvement in the distortion matrix would come from
its extension beyond $|\rpar|=300\hMpc$ to improve
the modeling of the cross-correlation beyond the BAO peak.
It would also be interesting to use narrower $(\rperp,\rpar)$ bins and independent redshift bins, and  to include the higher order terms
from eqn. \ref{eq:dhatexample}, in the calculation of the matrix.

There are several possibilities for further improvements
of the tests of the distortion matrix.
It would be good to repeat these tests with mocks that include
astrophysical contamination from metals and DLAs.
Instrumental effects like quasar redshift errors deserve more study.
Large redshift errors can strongly affect the cross-correlations
at small scales \citep{Bault_2025JCAP...01..130B}
and at large scales \citep{Youles_2022MNRAS.516..421Y}
unless care is taken to remove quasar-pixel pairs
that are sensitive to this effect \citep{Gordon_2025arXiv250508789G}.
The modeling of the redshift-error effect has been studied \citep{Gordon_2025arXiv250508789G}
but further verification that we understand the effect of distortions on the model would be useful.
Finally, more precise measurements of the auto-correlation redshift-evolution parameter, 
$\gamma_\alpha$, are necessary
to quantify this source of uncertainty in the distortion matrix.

\section{Data availability}
The data points corresponding to each figure in this paper can be accessed in the Zenodo
repository at 
\href{https://doi.org/10.5281/zenodo.15671861}{https://doi.org/10.5281/zenodo.15671861}.

\acknowledgments

 This material is based upon work supported by the U.S. Department of Energy (DOE), Office of Science, Office of High-Energy Physics, under Contract No. DE–AC02–05CH11231, and by the National Energy Research Scientific Computing Center, a DOE Office of Science User Facility under the same contract. Additional support for DESI was provided by the U.S. National Science Foundation (NSF), Division of Astronomical Sciences under Contract No. AST-0950945 to the NSF’s National Optical-Infrared Astronomy Research Laboratory; the Science and Technology Facilities Council of the United Kingdom; the Gordon and Betty Moore Foundation; the Heising-Simons Foundation; the French Alternative Energies and Atomic Energy Commission (CEA); the National Council of Humanities, Science and Technology of Mexico (CONAHCYT); the Ministry of Science, Innovation and Universities of Spain (MICIU/AEI/10.13039/501100011033), and by the DESI Member Institutions:
 https://www.desi.lbl.gov/collaborating-institutions . 
 Any opinions, findings, and conclusions or recommendations expressed in this material are those of the author(s) and do not necessarily reflect the views of the U. S. National Science Foundation, the U. S. Department of Energy, or any of the listed funding agencies.

AFR is partially supported by the European Union’s Horizon Europe research and innovation programme (COSMO-LYA, grant agreement 101044612), by the Spanish Ministry of Science and Innovation under the Ramon y Cajal program (RYC-2018-025210) and the PGC2021-123012NB-C41 project. IFAE is partially funded by the CERCA program of the Generalitat de Catalunya.

The authors are honored to be permitted to conduct scientific research on I'oligam Du'ag (Kitt Peak), a mountain with particular significance to the Tohono O’odham Nation.

\appendix

\section{The structure of the distortion matrix}
\label{sec:structure}

The distortion matrices $D^{auto}_{AA^\prime}$ and $D^{cross}_{AA^\prime}$ 
defined by eqns. \ref{eq:dmauto} and \ref{eq:dmcross} have
an index $A$ referring to a bin of the measured correlation function and an 
index $A^\prime$ referring to a bin of the true correlation function.
The indices $A$ and $A^\prime$ are, in turn, each defined by two indices, $(irt,irp)$,
defining ranges in $(\rperp,\rpar)$.
The bins of $A$ are of width $4\hMpc$ in both directions covering the range 
$0<\rperp<200\hMpc$ and $0<\rpar<200\hMpc$ (auto-corrrelation) and
$0<\rperp<200\hMpc$ and $-200<\rpar<200\hMpc$ (cross-correlation).
The $4\hMpc$ wide bins of $A^\prime$ for this study cover an extended $\rpar$ range:
$0<\rpar<300\hMpc$ (auto-corrrelation) and
$-300<\rpar<300\hMpc$ (cross-correlation).
The BOSS-eBOSS studies used maximum values of $|\rpar|$ of $200\hMpc$ for both $A$ and $A^\prime$.\footnote{While the $A$ bins that are used cover only the range $0<|\rpar|<200\hMpc$, \texttt{picca} calculates the distortion matrix over the $A$ range $0<|\rpar|<300\hMpc$ (the same as the $A^\prime$ range) so that the distortion matrix is square.}

\begin{figure}
    \centering
    \includegraphics[width=0.49\textwidth]{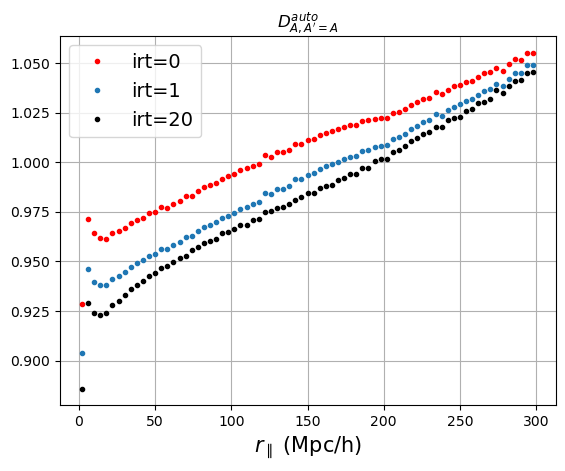}
    \includegraphics[width=0.49\textwidth]{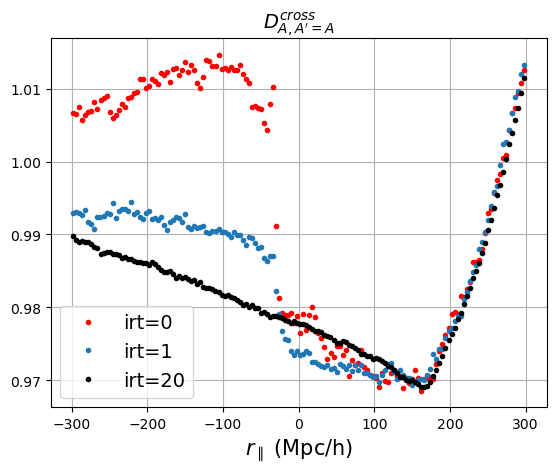}
    \caption{Some diagonal elements of the distortion matrices as a function of $\rpar$ for selected values of $\rperp$ (defined by $irt$ described in the text).}
    \label{fig:xdm_cf} 
\end{figure}

\begin{figure}
    \centering
    \includegraphics[width=0.47\textwidth]{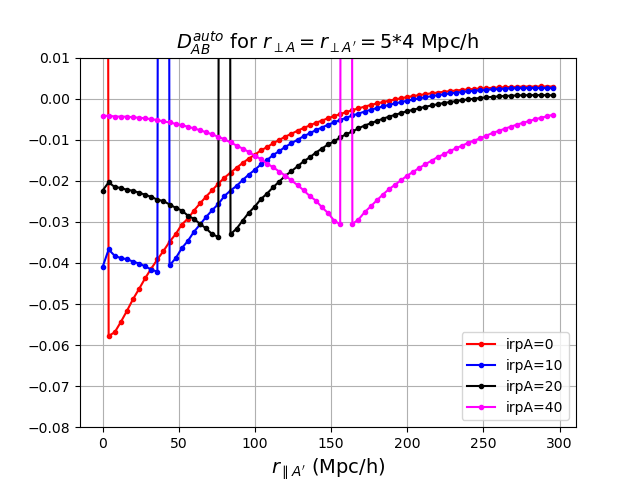}
    \includegraphics[width=0.47\textwidth]{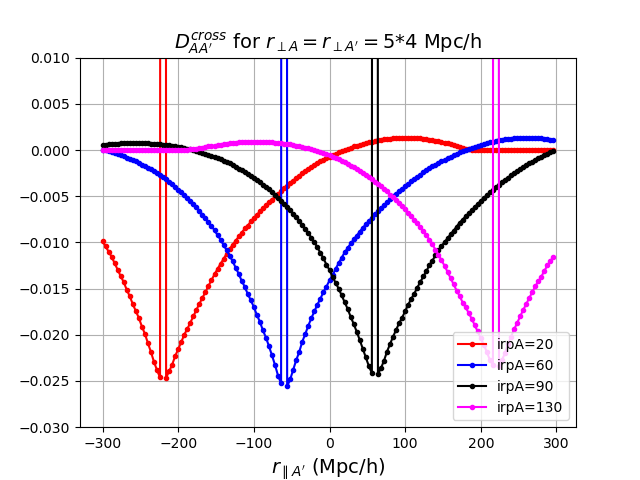}
    \caption{Some off-diagonal elements of the distortion matrices as a function of $r_{\parallel A^\prime}$ for $r_{\perp A}=r_{\perp A^\prime}=20\hMpc$  for selected values of $r_{\parallel A}$ (defined by $irp$ described in the text). The diagonal elements ($\approx 1$) are off-scale.}
    \label{fig:dm_cf}
\end{figure}

Examples of the distortion matrix elements are shown in Figs. \ref{fig:xdm_cf} and \ref{fig:dm_cf}.
As expected, the diagonal elements are of order unity and the off-diagonal elements are a few percent.

\begin{figure}
    \centering
    \includegraphics[width=0.48\textwidth]{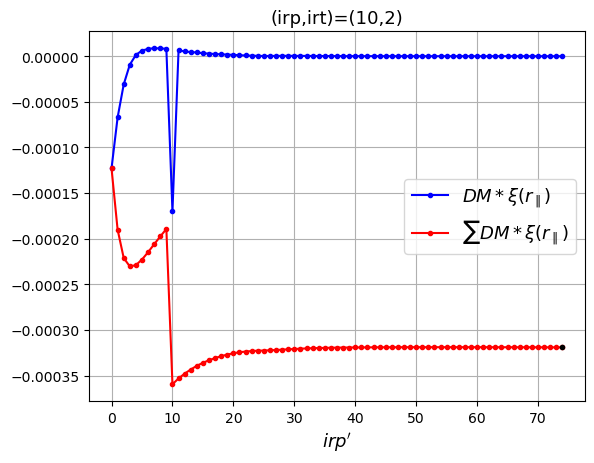}
    \includegraphics[width=0.45\textwidth]{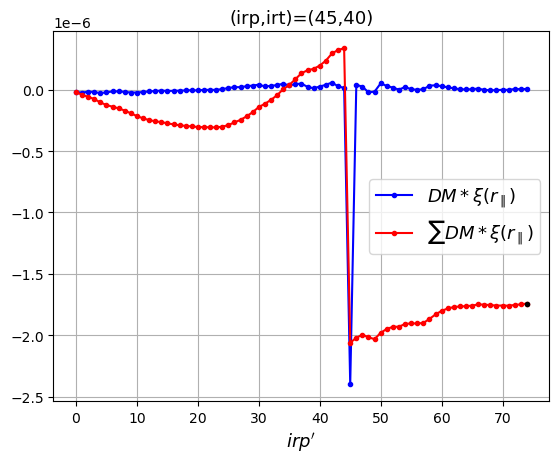}
    \caption{The construction of the distorted auto-correlation function for two $(\rperp,\rpar)$  indexed by $(irp,irt)$
    bins showing (blue) the contribution for each $\rpar^\prime$ value and (red) the sum of the contributions up to a given $\rpar^\prime$} 
    \label{fig:cfint}
\end{figure}

The construction of the distorted correlation function from the true correlation function is illustrated in Fig. \ref{fig:cfint}.
Two $(\rperp,\rpar)$ bins are shown:  
on the left the $(irp,irt)=(10,2)$ bin corresponding to $(\rpar,\rperp)=(40,8)\hMpc$ and
on the right the $(irp,irt)=(45,40)$ bin corresponding to $(\rpar,\rperp)=(180,160)\hMpc$. 
For each $(\rperp\rpar)$ bin, the final distorted correlation is the sum over correlations of different $(\rperp^\prime,\rpar^\prime)$ weighted by the matrix element. Because forest lines of sight are nearly parallel,
the distortion matrix is non-zero only for 
$\rperp^\prime\approx\rperp$.  The blue points show the contribution for each $\rpar^\prime$ and the red points show the sum up to a given $\rpar^\prime$.

\begin{figure}
    \centering
    \includegraphics[width=0.49\textwidth]{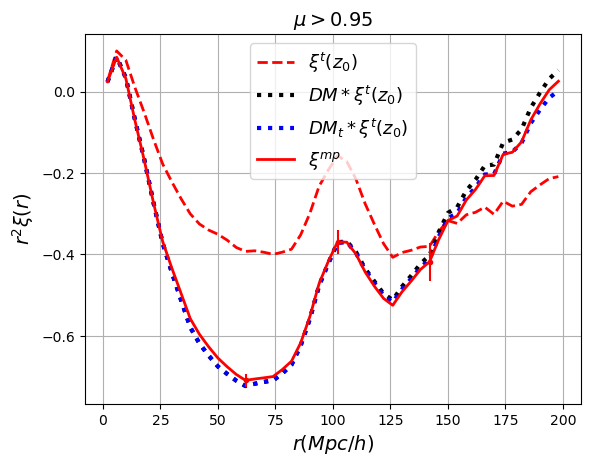}
    \includegraphics[width=0.49\textwidth]{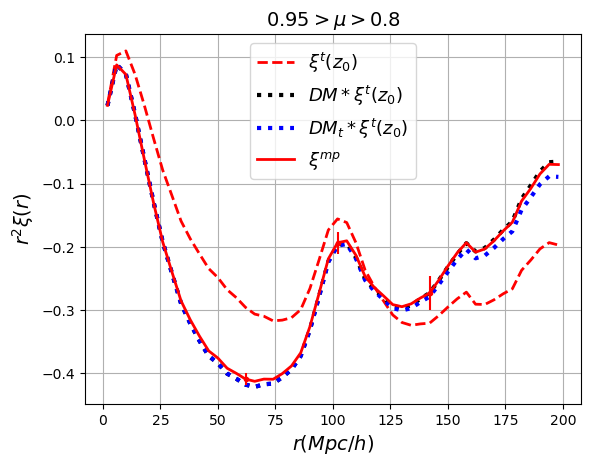}\\
    \includegraphics[width=0.49\textwidth]{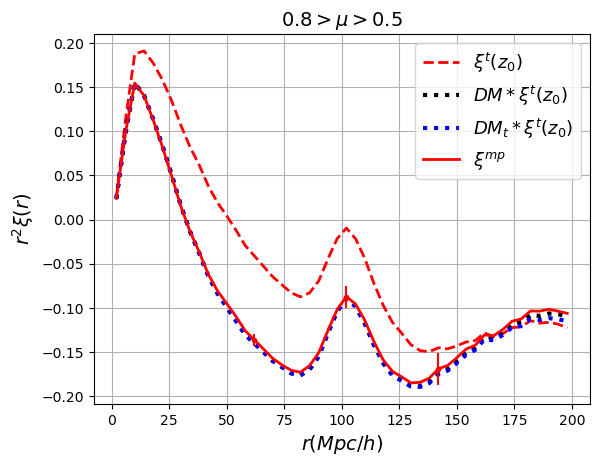}
    \includegraphics[width=0.49\textwidth]{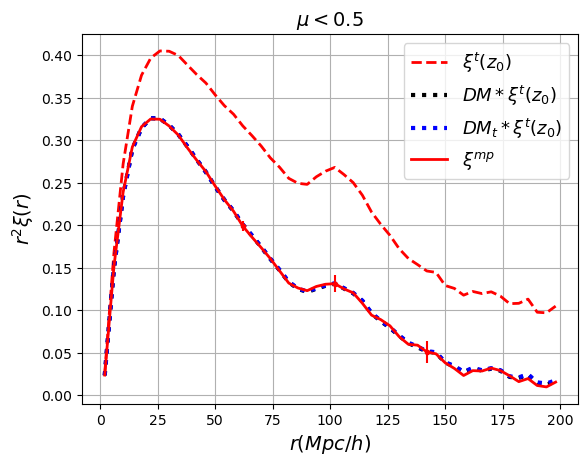}\\    
    \caption{Same as Fig. \ref{fig:zcfwedges} except that, in addition to the black dotted
    lines showing the true correlation function distorted by the distortion matrix, the blue
    dotted lines show the same function except using a distortion matrix extending only to 
    $\rpar=200\hMpc$.}
    \label{fig:cftrunc} 
\end{figure}

The $(10,2)$ point at low $\rperp$ receives major contributions from itself and, as expected, 
contributions from the large correlations at low $\rpar^\prime$.
There is little contribution from large $\rpar^\prime$ so the summed correlation quickly reaches its asymptotic value by $irp^\prime\approx20$.

The situation is quite different for the (45,40) bin at high $\rperp$  and $\rpar$.
There is still the major self-contribution but the remaining contribution are spread
over a wide $\rpar^\prime$ range.  The summed correlation function only reaches 
the asymptotic value around $\rpar^\prime\approx240\hMpc$.  This result was the reason that
the DESI distortion matrix was extended to $300\hMpc$ compared to the $200\hMpc$  for BOSS and eBOSS.  

The change from $200$ to $300\hMpc$ has a noticeable effect on the  distorted correlations above $150\hMpc$,
as can be seen in Fig. \ref{fig:cftrunc}.  Here the black dotted lines use the $300\hMpc$ distortion matrix and the blue dotted lines use the $200\hMpc$ matrix.  
Below $r=150\hMpc$ the blue and black dots are superimposed and then separate at higher values of $r$.

\begin{figure}
    \centering
    \includegraphics[width=0.48\textwidth]{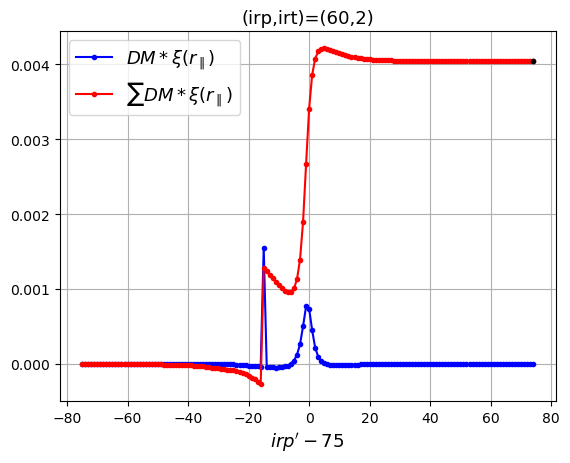}
    \includegraphics[width=0.46\textwidth]{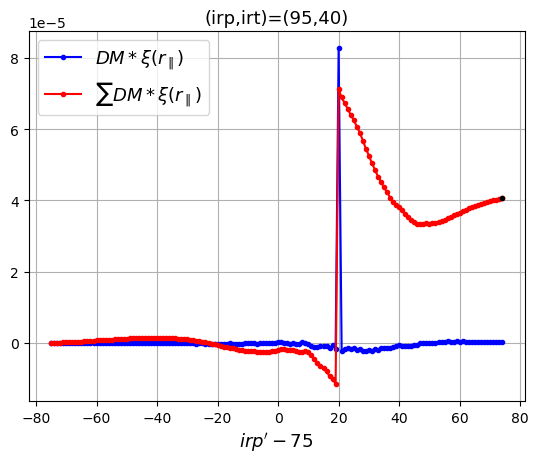}
    \caption{
    The construction of the distorted cross-correlation function for two $(\rperp,\rpar)$ indexed by $(irt,irp)$
    bins showing (blue) the contribution for each $\rpar^\prime$ value and (red) the sum of the contributions up to a given $\rpar^\prime$
    }
    \label{fig:xcfint}
\end{figure}

The construction of the cross-correlation function, illustrated in Fig. \ref{fig:xcfint}, has the same features as the auto-correlation function.  The low $(\rpar,\rperp)$ bin, $(irp,irt=(60,2)$
corresponding to $(\rpar,\rperp)=(-60,8)\hMpc$ has a large self-contribution and a contribution from bins at $\rpar\approx0$.  The summed contribution rapidly reaches its asymptotic value.

The large $(\rpar,\rperp)$ bin $(irp,irt)=(95,40)$ corresponding to $(\rpar,\rperp)=(80,160)\hMpc$
has a large self-contribution and a spread out off-diagonal contribution.  The summed cross-correlation reaches its asymptotic value only at $\rpar\approx280\hMpc$.
Higher values of $\rpar$ do not reach an asymptote indicating that the $300\hMpc$ extent
of the distortion matrix is insufficient, consistent with  the discrepancies at large $\rpar$ seen in Fig. \ref{fig:zcxfwedges} for $|\rpar|>100\hMpc$.

\begin{figure}
    \centering
    \includegraphics[width=0.47\textwidth]{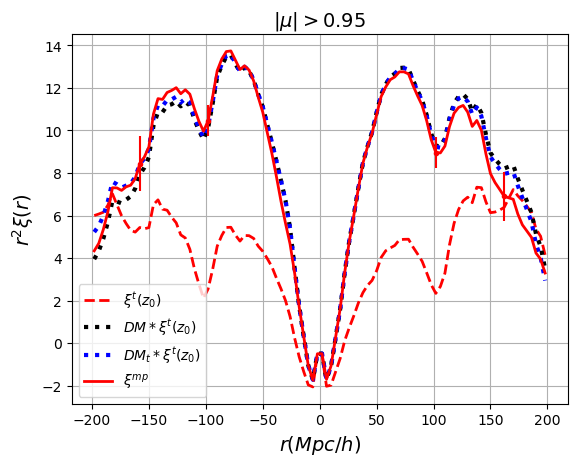}
    \includegraphics[width=0.47\textwidth]{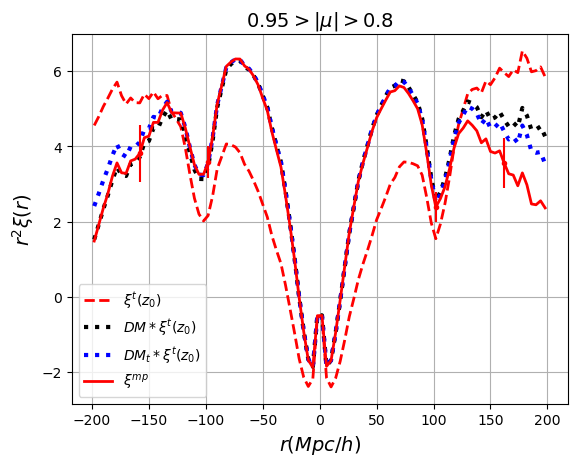}\\
    \includegraphics[width=0.47\textwidth]{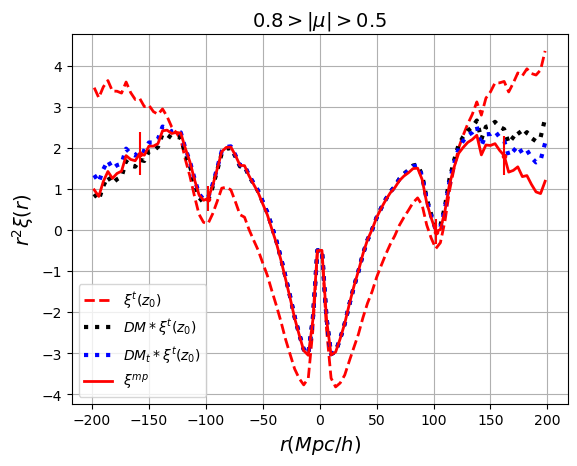}
    \includegraphics[width=0.47\textwidth]{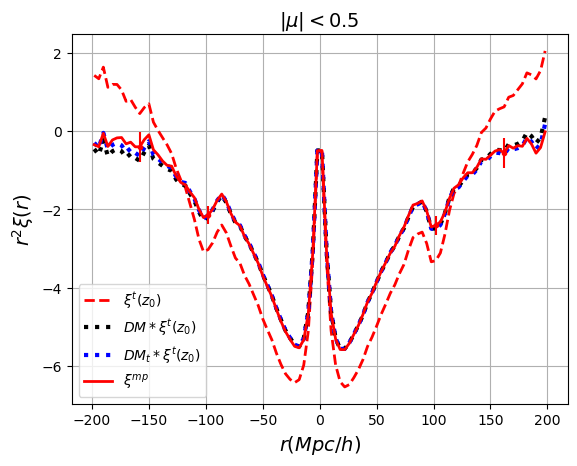}\\    
    \caption{
    Same as Fig. \ref{fig:cftrunc} except now for the cross-correlation function.
    }
    \label{fig:xcftrunc} 
\end{figure}

As an illustration of the changes that can be expected by expanding the distortion matrix,
Figure \ref{fig:xcftrunc} shows the changes in the cross-correlation function when expanding the distortion matrix from $200$ to $300\hMpc$.

\bibliographystyle{JHEP.bst}
\bibliography{references}
\appendix

\end{document}